\documentclass[aps,pre,twocolumn,superscriptaddress]{revtex4}

\usepackage{amsmath}
\usepackage{amssymb}
\usepackage{amsfonts}
\usepackage{amsthm}
\usepackage[dvips]{graphicx}
\DeclareGraphicsExtensions{.eps}

\begin{document}
\title{Three applications of path integrals: equilibrium and
kinetic isotope effects, and the temperature dependence of the rate constant 
of the [1,5] sigmatropic hydrogen shift in (Z)-1,3-pentadiene}
\author{Tom\'{a}\v{s} Zimmermann}
\author{Ji\v{r}\'{\i} Van{\'{\i}}{\v{c}}ek}
\email{jiri.vanicek@epfl.ch}
\affiliation{Laboratory of Theoretical Physical Chemistry, Institut des Sciences et
Ing\'{e}nierie Chimiques, \'{E}cole Polytechnique F\'{e}d\'{e}rale de Lausanne
(EPFL), CH-1015 Lausanne, Switzerland}
\date{\today}

\begin{abstract}
Recent experiments have confirmed the importance of nuclear quantum effects
even in large biomolecules at physiological temperature. Here we describe how
the path integral formalism can be used to describe rigorously the nuclear
quantum effects on equilibrium and kinetic properties of molecules.
Specifically, we explain how path integrals can be employed to evaluate the
equilibrium (EIE) and kinetic (KIE) isotope effects, and the temperature
dependence of the rate constant. The methodology is applied to the [1,5]
sigmatropic hydrogen shift in pentadiene. Both the KIE and the temperature
dependence of the rate constant confirm the importance of
tunneling and other nuclear quantum effects as well as of the anharmonicity 
of the potential energy surface. Moreover, previous results on
the KIE were improved by using a combination of a high level electronic
structure calculation within the harmonic approximation with a path integral
anharmonicity correction using a lower level method.

\end{abstract}
\maketitle


\section{Introduction}

After solving the electronic structure problem, most molecular modeling
simulations treat nuclei as classical particles. While this often is an
appropriate point of view, it fails completely in some cases, and almost
always when hydrogen is involved in bond breaking or formation. Strong nuclear
quantum effects have been observed in many reactions and recently even in several enzymatic reactions
\cite{kohen:1999,basran:1999,masgrau:2006}. In such situations, one should
treat some or all nuclei quantum mechanically. In particular, the nuclear quantum 
effects should be considered in the simulation even though they
may be similar for the corresponding reactions in enzyme and in solution, and 
therefore may not contribute to the catalytic effect of the enzyme \cite{Doll2003,Olsson2004}.

Solving the time-dependent Schr\"{o}dinger equation exactly is of course
possible only for a handful of atoms \cite{clary:2008}. Luckily, in most
chemical reactions, some quantum effects are at least partially washed out by
finite temperature. Whereas the \textquotedblleft real-time\textquotedblright%
\ quantum dynamics is extremely difficult, quantum thermodynamics, or
\textquotedblleft imaginary-time\textquotedblright\ quantum dynamics, can be
computed accurately for fairly large systems using the Feynman path
integrals \cite{Feynman1965,Kleinert2004}.

In general physics, path integrals are mostly\ known as an elegant tool to
formulate analytical theories in particle physics and quantum field theory. In
chemical physics, the somewhat abstract path integrals have evolved into an
extremely practical numerical tool \cite{Berne1986,ceperley:1995}. In fact,
they are probably the most successful tool in solving quantum statistical
problems in large systems without symmetry.

In this paper, we describe how path integrals can be used to compute the
equilibrium isotope effects, kinetic isotope effects, and the temperature
dependence of the reaction rate constant.

\section{Methodology}
\subsection{Path integral methods}

The central quantity in quantum statistical mechanics is the partition
function,%
\begin{equation}
Q(\beta)=\sum_{n}e^{-\beta\varepsilon_{n}} \label{q_def}%
\end{equation}
where $\beta=\left(  k_{B}T\right)  ^{-1}$ is the inverse temperature and
$\varepsilon_{n}$ is the energy of the system in the eigenstate $n$. If the
partition function is known analytically, any thermodynamic quantity can be
found. E.g., the thermal energy can be computed as%
\begin{equation}
E=-\frac{\partial\ln Q(\beta)}{\partial\beta}. \label{e_from_q}%
\end{equation}

The beauty of path integrals lies in that they allow computing the partition
function without finding the eigenstates of the Hamiltonian. Let us therefore
consider a molecular system consisting of $N$ atoms with masses $m_{i}$.
Starting from the exact expression $Q(\beta)=\operatorname{Tr}(e^{-\beta
\hat{H}})$, one obtains the path integral (PI) representation of $Q$ as
$Q=\lim_{P\rightarrow\infty}Q_{P}$, with \cite{ceperley:1995}%

\begin{equation}
Q_{P}(\beta)=C\int d\mathbf{r}^{\left(  1\right)  }\cdots\int d\mathbf{r}%
^{\left(  P\right)  }\exp\left[  -\beta\Phi\left(  \left\{  \mathbf{r}%
^{\left(  s\right)  }\right\}  \right)  \right]  . \label{q_PI}%
\end{equation}
Above, $C\equiv\left(  \frac{P}{2\pi\hbar^{2}\beta}\right)  ^{3NP/2}%
\prod_{i=1}^{N}m_{i}^{3P/2}$ is a multiplicative factor, $P$ the number of
imaginary time slices, and $\mathbf{r}^{\left(  s\right)  }\equiv
(\mathbf{r}_{1}^{\left(  s\right)  },\mathbf{r}_{2}^{\left(  s\right)
},\ldots,\mathbf{r}_{N}^{\left(  s\right)  })$ the set of Cartesian
coordinates associated with the $s$th time slice. Finally, $\Phi\left(
\left\{  \mathbf{r}^{\left(  s\right)  }\right\}  \right)  $ is the effective
potential\ given by%
\begin{eqnarray}
\Phi\left(  \left\{  \mathbf{r}^{\left(  s\right)  }\right\}  \right)
=\frac{P}{2\hbar^{2}\beta^{2}}\sum_{s=1}^{P}\sum_{i=1}^{N}m_{i}\left(
\mathbf{r}_{i}^{\left(  s\right)  }-\mathbf{r}_{i}^{\left(  s-1\right)
}\right)  ^{2}
\notag\\
+\frac{1}{P}\sum_{s=1}^{P}V\left(  \mathbf{r}^{\left(  s\right)
}\right)  \label{phi_PI}%
\end{eqnarray}
with $\mathbf{r}^{\left(  0\right)  }\equiv\mathbf{r}^{\left(  P\right)  }$
and $\left\{  \mathbf{r}^{\left(  s\right)  }\right\}  $ representing
$\left\{  \mathbf{r}^{\left(  1\right)  },\mathbf{r}^{\left(  2\right)
},\ldots,\mathbf{r}^{\left(  P\right)  }\right\}  $.

From expressions (\ref{q_PI}) and (\ref{phi_PI}) it is obvious that for $P=1$
one obtains the classical partition function and therefore classical
thermodynamics. The quantum thermodynamics is obtained in the limit
$P\rightarrow\infty$, but in practice it often suffices to take a finite value
of $P$ to obtain accurate quantum results.

To compute a thermal average $A(\beta)$ of a physical observable $A$ such as
the heat capacity or a rate constant, one starts from the exact quantum
expression $A(\beta)=\operatorname{Tr}[\hat{A}\exp(-\beta\hat{H})]/Q(\beta)$
or, if possible, from an exact expression for $A(\beta)$ in terms of the
partition function, such as Eq. (\ref{e_from_q}) for energy. Using the PI
expression (\ref{q_PI}) for $Q$, one ends up with a PI expression for the
thermal average $A(\beta)$,%
\begin{equation}
A(\beta)=\left\langle A\left(  \left\{  \mathbf{r}^{\left(  s\right)
}\right\}  \right)  \right\rangle _{\rho}, \label{pi_avg}%
\end{equation}
where $\left\langle \cdot\right\rangle _{\rho}$ denotes a weighted average
over PI\ integral configurations. The weight is given by $\rho=\exp(-\beta
\Phi)$ and the quantity $A\left(  \left\{  \mathbf{r}^{\left(  s\right)
}\right\}  \right)  $ is called a PI \textquotedblleft
estimator\textquotedblright\ for $A(\beta)$.

The PI average (\ref{pi_avg}) can be evaluated efficiently using a PI
molecular dynamics (PIMD) or PI\ Monte Carlo (PIMC) techniques
\cite{ceperley:1995}. In the calculations presented below, a PIMD
implementation in the molecular dynamics package AMBER 10 \cite{amber10} was
used for the equilibrium isotope effects while an in-house PIMC code was used
for the kinetic isotope effects and the temperature dependence of the rate constant.

Up to this point, everything was straightforward. The interesting twist comes
because the estimator $A\left(  \left\{  \mathbf{r}^{\left(  s\right)
}\right\}  \right)  $ is not a unique function of $\left\{  \mathbf{r}%
^{\left(  s\right)  }\right\}  $. The art of PIs lies in finding the optimal
estimator in the sense of having the smallest statistical error for a given
number $M$ of samples. As the statistical error of a simulation is
proportional to $M^{-1/2}$, an estimator with a smaller statistical error can
lead to a much more efficient simulation.

\subsection{Equilibrium isotope effects}

The first application that we shall consider is the calculation of the
equilibrium isotope effect (EIE), defined as the effect of isotopic
substitution on the equilibrium constant. More precisely,%
\begin{equation}
\text{EIE}=\frac{K_{l}}{K_{h}},\label{eie_def}%
\end{equation}
where $K$ is the equilibrium constant and $l$ ($h$) denotes the lighter
(heavier) isotope. The equilibrium constant can be expressed in terms of
partition functions as $K=Q_{p}/Q_{r}$ where subscripts $p$ and $r$ denote the product and reactant, respectively. Hence the EIE can be computed
as
\begin{equation}
\text{EIE}=\frac{\exp\left[  -\beta\int_{0}^{1}d\lambda\frac{dF_{p}%
(\lambda)}{d\lambda}\right]  }{\exp\left[  -\beta\int_{0}^{1}d\lambda
\frac{dF_{r}(\lambda)}{d\lambda}\right]  },\label{eie_TI}%
\end{equation}
where $F=-\beta^{-1}\log Q$ is the free energy and $\lambda$ is a parameter
providing a smooth transition between the lighter and heavier isotopologs
\cite{zimmermann:2009}. For instance, it can be accomplished by a linear
interpolation of masses of all atoms in a molecule according to the equation%
\begin{equation}
m_{i}(\lambda)=\left(  1-\lambda\right)  m_{h,i}+\lambda m_{l,i}%
.\label{mass_interpol}%
\end{equation}
Unlike the partition function or the equilibrium constant $K$, 
the derivative $dF/d\lambda$ of the free
energy can be computed by a PIMD or PIMC simulation directly, and so the
equilibrium constant can be computed by the so-called thermodynamic
integration (TI) expressed in Eq. (\ref{eie_TI}). The estimator for $dF/d\lambda$
can be found in Ref. \onlinecite{zimmermann:2009}.

\subsection{Quantum instanton approximation}

Theoretically more challenging quantities than the equilibrium constant or EIE
are the rate constant or the kinetic isotope effect (KIE). The reason is that
these kinetic quantities combine quantum thermodynamics with the real-time
quantum dynamics. However, since the most interesting chemical reactions occur
at a thermodynamically relatively high temperature, there exist suitable approximations.

A recent and accurate approximation which takes into account most quantum
effects is the so-called quantum instanton (QI) approximation of Miller and
coworkers \cite{miller:2003}. In this approximation, the rate constant is
expressed as%
\begin{equation}
k_{\text{QI}}(\beta)=\frac{\sqrt{\pi}}{2}\frac{\hbar}{\Delta H}\frac
{C_{\text{ff}}\left(  0\right)  }{Q_{r}}, \label{k_QI}%
\end{equation}
where $C_{\text{ff}}(t)$ is the flux-flux correlation function, $Q_{r}$ the
reactant partition function, and $\Delta H$ a certain energy variance near the
transition state.

\subsection{Kinetic isotope effects}

The kinetic isotope effect is defined as the effect of isotopic substitution
on the rate constant,%
\begin{equation}
\text{KIE}=\frac{k_{l}}{k_{h}}. \label{kie_def}%
\end{equation}
This KIE is widely used in chemical kinetics to detect nuclear tunneling and
other quantum effects, as well as to distinguish between possible reaction
mechanisms. Unlike the rate constant itself, the KIE depends very little on
the classical energy barrier height, and so the KIE can separate various
effects from the simple exponential dependence on the barrier that overwhelms
the rate constant. Using the QI expression (\ref{k_QI}), the KIE can be
expressed as \cite{vanicek:2005}%
\begin{equation}
\text{KIE}=\frac{Q_{r,h}}{Q_{r,l}}\frac{\Delta H_{h}}{\Delta H_{l}}%
\frac{C_{\text{dd,}l}(0)}{C_{\text{dd,}h}(0)}\frac{\frac{C_{\text{ff},l}%
(0)}{C_{\text{dd,}l}(0)}}{\frac{C_{\text{ff},h}(0)}{C_{\text{dd,}h}(0)}},
\label{kie_QI}%
\end{equation}
where the delta-delta correlation function $C_{\text{dd}}(t)$ was introduced.
This correlation function at time $t=0$ is a generalization of the partition
function, constrained to two dividing surfaces. Introduction of $C_{\text{dd}%
}$ into expression (\ref{kie_QI}) simplifies the calculation: On one hand,
similarly to $\Delta H$ the flux factor $C_{\text{ff}}(0)/C_{\text{dd}}(0)$
can be computed directly in a single PIMC or PIMD simulation
\cite{yamamoto:2004}. On the other hand, the ratio $C_{\text{dd,}%
l}(0)/C_{\text{dd,}h}(0)$ can be computed by a TI
analogous to Eq. (\ref{eie_TI}). The estimators can be found in Refs.
\onlinecite{vanicek:2005,vanicek:2005a,vanicek:2007}.

It should be noted that several other PI approaches to compute the
KIE exist, which are however not based on the QI. These include, e.g.,
approaches based on other quantum transition state theories
\cite{Calhoun1998,Schofield2002} or on the quantized classical path method \cite{Hwang1991,Olsson2004,Liu2007}.

\subsection{Temperature dependence of the rate constant}

The last application that we shall consider is the direct evaluation of the
temperature dependence of the rate constant, i.e., of the ratio $k(T)/k(T_{0}%
)$. Again, this quantity is extremely useful in kinetics because it can help
to discern between plausible mechanisms of a reaction and because the ratio is
easier to measure accurately than the rate constant itself.

In the framework of the QI\ approximation, the $T$ dependence can be evaluated
as \cite{buchowiecki:2010}%
\begin{equation}
\frac{k_{\text{QI}}(\beta)}{k_{\text{QI}}(\beta_{0})}=\frac{Q_{r}(\beta_{0}%
)}{Q_{r}(\beta)}\frac{\Delta H(\beta_{0})}{\Delta H(\beta)}\frac{C_{\text{dd}%
}\left(  \beta\right)  }{C_{\text{dd}}\left(  \beta_{0}\right)  }\frac
{\frac{C_{\text{ff}}\left(  \beta\right)  }{C_{\text{dd}}\left(  \beta\right)
}}{\frac{C_{\text{ff}}\left(  \beta_{0}\right)  }{C_{\text{dd}}\left(
\beta_{0}\right)  }}.\label{tdep_QI}%
\end{equation}
The flux and energy variance are computed as for the KIEs, but the ratios of
the partition functions and of the delta-delta correlation functions use a
different type of TI, namely a TI\ with respect to the inverse temperature
$\beta$ \cite{ceotto:2004,buchowiecki:2010}. Taking advantage of the relations%
\begin{align}
E_{r}(\beta) &  =-\frac{d\log Q_{r}(\beta)}{d\beta}\text{ and}%
\label{e_r_logder}\\
E^{\ddag}(\beta) &  =-\frac{d\log C_{\text{dd}}(\beta)}{d\beta}%
\label{e_TS_logder}%
\end{align}
between the reactant and transition state energies and the logarithmic derivatives of
$Q_{r}$ and $C_{\text{dd}}$, one can compute the ratios of $Q$s and
$C_{\text{dd}}$s as%
\begin{align}
\frac{Q_{r}(\beta)}{Q_{r}(\beta_{0})} &  =\exp\left[  -\int_{\beta_{0}}%
^{\beta}E_{r}(\tilde{\beta})d\tilde{\beta}\right]  ,\label{ti_qr}\\
\frac{C_{\text{dd}}\left(  \beta\right)  }{C_{\text{dd}}\left(  \beta
_{0}\right)  } &  =\exp\left[  -\int_{\beta_{0}}^{\beta}E^{\ddag}(\tilde
{\beta})d\tilde{\beta}\right]  .\label{ti_cdd}%
\end{align}
The estimators for the reactant energy $E_{r}$ are well-known for PIMD
simulations \cite{Barker1979,Herman1982,Parrinello1984}. Estimators for the transition state energy $E^{\ddag}$ were developed in Ref.~\onlinecite{buchowiecki:2010} for several types of constraints. These estimators are derived in a similar manner, based on the rescaling of coordinates and a finite difference evaluation of the derivative with respect to $\beta$ that was used for the first time by Predescu and coworkers for reactant energies and heat capacities \cite{Predescu2003b}. The statistical error of these estimators showed a surprising behavior in comparison with the estimators for $E_{r}$. In particular, the centroid virial estimators was not always the optimal estimator \cite{buchowiecki:2010}. Below we use
estimators from Ref. \onlinecite{buchowiecki:2010} for both $E_{r}$ and $E^{\ddag}$ because they are suitable for a PIMC simulation.

\section{Computational details}

\subsection{Equilibrium isotope effects}

To compute the EIE, ratios of partition functions corresponding to
different isotopologs of reactants and products have to be computed.
Unfortunately, higher level \textit{ab initio} electronic structure
methods are usually too expensive to be used in the PI calculation
directly. On the other hand, semiempirical or force field methods
generally do not achieve comparable accuracy. To take advantage of
both the accuracy of \textit{ab initio} methods and the rigor of PI
treatment of nuclear motion, we combine the value of the EIE obtained
in the harmonic approximation (HA) using a higher level method with
the anharmonicity correction, rigorously computed with the PIMD method,
using a lower level method. The anharmonicity correction is calculated
as \begin{equation}
\Delta\Delta F^{\mathrm{anharm}}=\Delta F_{\mathrm{PI}}^{\mathrm{red}}-\Delta F_{\mathrm{HA}}^{\mathrm{red}},\label{eq:F_anharm}\end{equation}
 where $\Delta F_{\mathrm{PI}}^{\mathrm{red}}$ and $\Delta F_{\mathrm{HA}}^{\mathrm{red}}$
are the free energy differences computed with the PI method and the
HA, respectively. The HA value of $\Delta F^{\mathrm{red}}$ is obtained
by Boltzmann averaging over all possible distinguishable conformations,
\begin{equation}
{\displaystyle \Delta F_{\mathrm{HA}}^{\mathrm{red}}=-k_{B}T\ln\left[\frac{s_{r}\sum_{i=1}^{N_{p}}\left\{ \exp\left(\frac{-E_{i}^{\mathrm{el}}}{k_{B}T}\right)\sum_{j=1}^{s_{p}}Q_{p,ij}^{\mathrm{nuc}}\right\} }{s_{p}\sum_{i=1}^{N_{r}}\left\{ \exp\left(\frac{-E_{i}^{\mathrm{el}}}{k_{B}T}\right)\sum_{j=1}^{s_{r}}Q_{r,ij}^{\mathrm{nuc}}\right\} }\right]},\label{eq:F_red_HA}\end{equation}
 where $N_{r}$ is the number of {}``geometrically different isomers''
of the reactant. By geometrically different isomers we mean
species differing in their geometry, not species differing only in
positions of isotopically substituted atoms. $E_{i}^{\mathrm{el}}$
is the electronic energy (including nuclear repulsion) of the $i^{\mathrm{th}}$
isomer, $s_{r}$ is the symmetry factor, which can arise from the
change of the wave function due to the effects of indistinguishability
of particles during isotopic substitution. Finally, $Q_{r,ij}^{\mathrm{nuc}}$
are partition functions of the nuclear motion of $s_{r}$ isotopomers.
$N_{p}$, $s_{p}$, $Q_{p,ij}^{\mathrm{nuc}}$ denote analogous
quantities for the product.

Since in our approximation the electronic function does not change
after isotopic substitution, the EIE is dominated by vibrational
contributions. Therefore, we searched for an optimal electronic structure method 
among those studied by Merrick \textit{et al.} \cite{Merrick2007},
who tested the performance of several higher level
methods by comparison of the computed vibrational frequencies in the
HA with experimental data for a set of 39 molecules. The B98/6-311+(2df,p)
method, which we chose, has the root mean square error (RMSE) of the zero point energy (ZPE)
equal to 0.31 $\mathrm{kJ\cdot mol^{-1}}$ and RMSEs of frequencies
31 $\mathrm{cm^{-1}}.$ The lower level method used to compute the PI anharmonicity
corrections was the AM1 semiempirical method \cite{Dewar1985},
which, in the HA, reproduces the B98/6-311+(2df,p) results very well.

Since the value of the EIE is expected to be close to unity, we have
chosen a relatively high number of time slices $P=40$ in the discretized
path integral. The TI was performed with Simpson's rule using five
values of $\lambda$. As the dependence of the PIMD average of $d F / d \lambda$
is almost linear over the full range of $\lambda$, five points were sufficient 
to get a converged result for
the integral. For further details about the method, see Ref. \onlinecite{zimmermann:2009}.

\subsection{Kinetic isotope effects}

The ratio of partition functions needed for the KIE was
computed in the same manner as the ratio for the EIE, with the exception
that the number of imaginary time slices was set to $P=24.$ Other terms
in the Eq.~(\ref{kie_QI}) were computed with the PIMC method
using the empirical valence bond (EVB) potential \cite{Warshel1980,Warshel1991}, which allows the molecular mechanics potential
to be fitted to match the B98/6-311+(2df,p) reaction barrier and Hessian
at the transition state \cite{Chang1990}.

The EVB potential is computed as 
\begin{equation}
V_{\text{EVB}}=\frac{1}{2}\left(V_{11}+V_{22}\right)-\sqrt{\left\vert V_{12}\right\vert ^{2}+\left(\frac{V_{11}-V_{22}}{2}\right)^{2}}
\end{equation}
where $V_{11}$ and $V_{22}$ are the molecular mechanics potential
energies of the reactant and product, respectively. These are
the diabatic potentials and the diagonal terms in the symmetric $2\times2$
EVB\ matrix. In our calculations, they were obtained with the general AMBER force field
(GAFF) \cite{wang:2004a}. The off-diagonal term
$V_{12}$, i.e., the coupling between the two diabatic states, was
calculated according to the formula of Schlegel and
Sonnenberg \cite{schlegel:2006}: 
\begin{equation}
V_{12}(\mathbf{r})=A[1+\mathbf{B}\cdot\Delta\mathbf{r}+\Delta\mathbf{r}\cdot(\mathbf{C}+\alpha\mathbf{I})\cdot\Delta\mathbf{r}]\exp(-\alpha|\Delta\mathbf{r}|^{2}/2).\label{evb_schl_sonn}\end{equation}
 The constants $A$, $\mathbf{B}$, and $\mathbf{C}$ are chosen to match
the barrier height and the Hessian of the \textit{ab initio} potential
at the transition state.

\subsection{Temperature dependence of the rate constant}

A similar procedure as for the KIE was used to compute the temperature
dependence of the rate constant. Since the major
contribution to the temperature dependence of the rate constant is
due to the reaction barrier height, we computed the barrier height
also by the single point coupled clusters CCSD(T)/cc-pVTZ and CCSD/aug-cc-pVTZ calculations
on B98/6-311+(2df,p) geometries.

In the PI calculation of the
temperature dependence of the rate constant,
several approaches were used to compute the ratio of reactant partition
functions at different temperatures. In the first approach,
denoted ``QI GAFF(PIMC)'', the ratio of the reactant partition
functions was computed directly in the PIMC simulation with the GAFF
force field. This could possibly cause a difficulty because
in the EVB method, the Hessian of potential energy surface at
the transition state is fitted to the much more accurate 
B98/6-311+(2df,p) Hessian. At
this point, the fitted surface has B98/6-311+(2df,p) properties in the HA,
with the anharmonicity given by a nonlinear combination of the GAFF force
fields describing the reactant and product. Therefore, a systematic 
error could arise due
to the different properties of the B98 and GAFF surfaces in the HA.
Fortunately, the harmonic vibrational free energy computed using the GAFF force
field is only about $0.5$ $\mathrm{kcal/mol}$ lower than the
B98/6-311+(2df,p) value.  Consequently, the results obtained with the QI GAFF(PIMC)
method are almost equal to the {}``QI B98(HA) GAFF(PIMC) anharm.''
results, obtained by augmenting the B98/6-311+(2df,p) free
energies in the HA with the anharmonicity PIMC correction computed
with the GAFF force field. The {}``QI B98(HA)'' method uses only
plain B98/6-311+(2df,p) free energies in the HA. Finally, probably the most accurate ``QI
B98(HA) AM1(PIMC) anharm.'' method uses the B98/6-311+(2df,p) free
energies in the HA with the anharmonicity PIMC correction computed
 with the AM1 semiempirical potential.

\subsection*{Used software}

All PIMD calculations were performed in Amber 10 \cite{amber10}. The
\textit{ab initio} and B98 calculations as well as AM1 semiempirical
calculations in the HA were done in Gaussian 03 revision E01 \cite{g03}.
All PIMC calculations were done using a PIMC code developed by one of
us.

\section{Results}

\begin{figure}
\includegraphics[width=\linewidth]{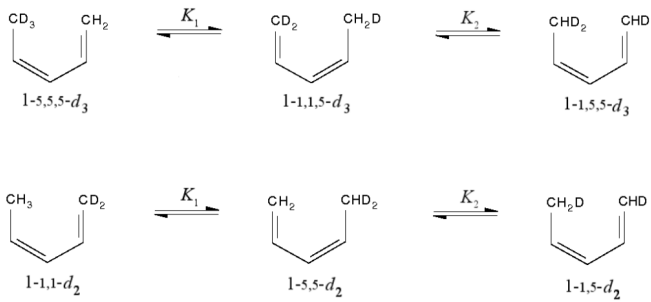}
\caption{The {[}1,5] hydrogen shift reaction in (3\textit{Z})-(5,5,5-$^{\text{2}}$H$_{\text{3}}$)penta-1,3-diene
(\textbf{1-}\textbf{\footnotesize 5,5,5-}\textbf{\textit{\footnotesize d}}\textbf{$_{\text{3}}$})
and in (3\textit{Z})-(1,1-$^{\text{2}}$H$_{2}$)penta-1,3-diene (\textbf{1-}\textbf{\footnotesize 1,1-}\textbf{\textit{\footnotesize d}}\textbf{\footnotesize $_{\text{2}}$}).
If all contributions except for those due to symmetry factors were
neglected, one would obtain approximate equilibrium constants $K_{1}=3$ for the
first reaction step and $K_{2}=2$ for the second step
 (in both cases).{\footnotesize \label{fig:equlibrium_pent}}}
\end{figure}

In this section, the path integral formalism is applied to study the
[1,5] sigmatropic hydrogen shift reaction in the (3\textit{Z})-penta-1,3-diene.
Two isotopologs of (3\textit{Z})-penta-1,3-diene are considered: tri-deuterated
(3\textit{Z})-(5,5,5-$^{\text{2}}$H$_{\text{3}}$)penta-1,3-diene
(\textbf{1-}\textbf{\footnotesize 5,5,5-}\textbf{\textit{\footnotesize d}}\textbf{\footnotesize $_{\text{3}}$}{\footnotesize )}
and di-deuterated (3\textit{Z})-(1,1-$^{\text{2}}$H$_{2}$)penta-1,3-diene
(\textbf{1-}\textbf{\footnotesize 1,1-}\textbf{\textit{\footnotesize d}}\textbf{\footnotesize $_{\text{2}}$})
(see Fig. \ref{fig:equlibrium_pent}). Both isotopologs were used
by Roth and K\"{o}nig in their experimental study of the KIE in the {[}1,5]
sigmatropic hydrogen shift reaction \cite{Roth1966}. The value of
the KIE obtained by Roth and K\"{o}nig was too high to be explained classically
implying that quantum effects such as tunneling are important in this
reaction. Here, we examine both the EIE and KIE for this reaction
as well as the temperature dependence of the rate constant in the
range studied in the experimental work \cite{Roth1966}.

\subsection{Equilibrium isotope effects}

The EIE for both steps of the {[}1,5] sigmatropic hydrogen shift reaction
is computed according to the Eq. (\ref{eie_TI}) and listed in Table
\ref{tab:EIE1}. We have already computed the equilibrium ratios for
both isotopologs in Ref. \onlinecite{zimmermann:2009}. For the convenience
of the reader, Table \ref{tab:EIE1} also shows these ratios.

The precision of the experimental ratios of Roth and K\"{o}nig is only
to a single significant digit, so it reflects only the ratios caused by the
symmetry effects and does not allow for a rigorous comparison with
the computed EIE. Nevertheless, in Ref. \onlinecite{zimmermann:2009}
we were able to use the same methodology to compare the theoretical
and experimental equilibrium ratios for a related compound (2,4,6,7,9-pentamethyl-5-methylene-11,11a-dihydro-12\textit{H}-naphthacene).
As the experimental data of Doering \textit{et al.} \cite{Doering2007}
were more accurate, a more rigorous test was possible, and led to
the conclusion that the current methodology can reproduce quantitatively
the equilibrium ratios beyond the symmetry effects. The value of the
EIE for the first reaction step is greater than unity, meaning that
the equilibrium of the first step reaction is shifted toward the product
side for the lighter isotopolog \textbf{1-}\textbf{\footnotesize 1,1-}\textbf{\textit{\footnotesize d}}\textbf{\footnotesize $_{\text{2}}$}
(where hydrogen is transferred) in comparison with the heavier \textbf{1-}\textbf{\footnotesize 5,5,5-}\textbf{\textit{\footnotesize d}}\textbf{\footnotesize $_{\text{3}}$}
(where the transferred atom is deuterium). In the second step, the
EIE is smaller than unity so the products are favored for the heavier
isotopolog. Also, in contrast to the first step, hydrogen is transferred
in the heavier isotopolog and deuterium in the lighter one.

\begin{table*}
\begin{tabular}{lccc}
\hline
EIE 
&
&
EIE\textit{$_{\text{1}}$= K}$_{\text{1}}$(\textbf{\textit{d}}\textbf{$_{\text{2}}$)/}\textit{K}$_{\text{1}}$(\textbf{\textit{d}}\textbf{$_{\text{3}}$)} &
EIE\textit{$_{\text{2}}$=K}$_{\text{2}}$(\textbf{\textit{d}}\textbf{$_{\text{2}}$)/}\textit{K}$_{\text{2}}$(\textbf{\textit{d}}\textbf{$_{\text{3}}$)}\tabularnewline
\hline 
AM1 (PIMD) &
&
1.07 &
0.96\tabularnewline
B98 (HA) +AM1 (PIMD) anharm. &
&
1.12 &
0.95\tabularnewline
\hline
Equilibrium fraction
&
\textbf{1-}\textbf{\footnotesize 5,5,5-}\textbf{\textit{\footnotesize d}}\textbf{\footnotesize $_{\text{3}}$} {\footnotesize }&
\textbf{1-}\textbf{\footnotesize 1,1,5-}\textbf{\textit{\footnotesize d}}\textbf{\footnotesize $_{\text{3}}$} {\footnotesize }&
\textbf{1-}\textbf{\footnotesize 1,5,5-}\textbf{\textit{\footnotesize d}}\textbf{\footnotesize $_{\text{3}}$}\tabularnewline
\hline
AM1 (PIMD) &
0.103 &
0.296 &
0.601\tabularnewline
\hline
B98 (HA) + AM1 (PIMD) anharm. &
0.104 &
0.293 &
0.602\tabularnewline
\hline
Equilibrium fraction
&
\textbf{1-}\textbf{\footnotesize 1,1-}\textbf{\textit{\footnotesize d}}\textbf{\footnotesize $_{\text{2}}$} {\footnotesize }&
\textbf{1-}\textbf{\footnotesize 5,5-}\textbf{\textit{\footnotesize d}}\textbf{\footnotesize $_{\text{2}}$} {\footnotesize }&
\textbf{1-}\textbf{\footnotesize 1,5-}\textbf{\textit{\footnotesize d}}\textbf{\footnotesize $_{\text{2}}$}\tabularnewline
\hline
AM1 (PIMD) &
0.099 &
0.305 &
0.597\tabularnewline
\hline
B98 (HA) +AM1 (PIMD) anharm. &
0.097 &
0.307 &
0.596\tabularnewline
\hline
\end{tabular}

\caption{
Values of the EIE on the first (\textit{K}$_{\text{1}}$) and second
(\textit{K}$_{\text{2}}$) step of the reactions from Fig. \ref{fig:equlibrium_pent} at 478.45 K. 
Results denoted
by AM1 (PIMD) were obtained with a PIMD calculation using the AM1
semiempirical method. Results denoted by B98 (HA) + AM1 (PIMD)
anharm. were obtained using the B98 method in the harmonic approximation
together with the anharmonicity correction computed with the AM1 method.
For reference, the bottom six rows show the equilibrium fractions of all isotopomers occurring 
in the reactions from Fig. \ref{fig:equlibrium_pent}. 
{\footnotesize \label{tab:EIE1}}}
\end{table*}

\subsection{Kinetic isotope effects}

\begin{figure}
\includegraphics[width=\linewidth]{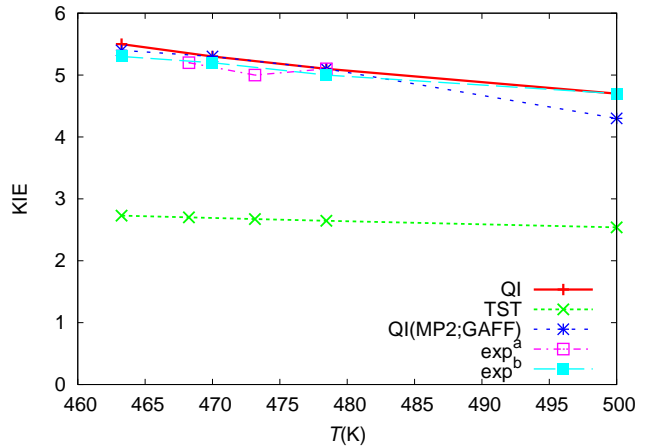}

\caption{The KIE on the first {[}1,5] hydrogen shift reaction from Fig.~\ref{fig:equlibrium_pent}. 
The displayed KIE is defined precisely in Eq. (\ref{our_KIE}) of the text. The curve marked as exp$^{\text{a}}$ denotes the raw experimental data from Ref. \onlinecite{Roth1966} whereas exp$^{\text{b}}$ denotes the data computed from an Arrhenius fit in Ref. \onlinecite{Roth1966}.
\label{fig:KIE}}
\end{figure}

Computed values of the KIE on the first reaction step are shown in
Figure \ref{fig:KIE} and Table \ref{tab:KIE}. This KIE is defined
as the ratio
\begin{equation}
\mathrm{KIE}=\frac{k(\textbf{1-}{\footnotesize \textbf{1,1-\textit{d}}\mathbf{_2}} \rightarrow \textbf{1-}{\footnotesize \textbf{5,5-\textit{d}}\mathbf{_2}})}
                  {k(\textbf{1-}{\footnotesize \textbf{5,5,5-\textit{d}}\mathbf{_3}} \rightarrow \textbf{1-}{\footnotesize \textbf{1,1,5-\textit{d}}\mathbf{_3}} )}.
\label{our_KIE}
\end{equation}
As can be seen from the table, the calculated and experimental data
agree very well; the largest relative difference is around 5 \%. Comparison
with the transition state theory (TST) values, computed using the B98/6-311+(2df,p) barrier
height and the partition functions in the HA [for reactants computed
according to Eq. (\ref{eq:F_red_HA})], clearly demonstrates the importance
of quantum effects on the KIE. (Note that in TST the value of the
KIE is actually independent of the barrier height). The third column
of the table contains the KIE computed in Ref. \onlinecite{vanicek:2007},
where the same PI method was used, but the barrier height and the
Hessian of the transition state were fitted to MP2/6-31g(d) ab initio
data. In addition, the GAFF force field \cite{wang:2004a} was used not only for
the EVB potential but also for the reactant partition functions. Our
results and results from Ref. \onlinecite{vanicek:2007} differ
by at most 10\% of the KIE value. The main improvement in comparison 
with Ref. \onlinecite{vanicek:2007} occurs at higher temperatures. 
However, the observed difference is not only due to
the use of different methods, but partially also due to the way the EVB potential
surface is constructed. As discussed in Ref. \onlinecite{vanicek:2007},
the results are weakly dependent on the value of $\alpha$ in Eq.
\eqref{evb_schl_sonn}, and even though a flat plateau is obtained
for a broad range of $\alpha$ values, varying $\alpha$ still can
change the KIE by about 10\%. Generally, the optimal value of
the parameter $\alpha$ can be different for different \textit{ab
initio} methods. Here we have intentionally chosen the same value
of $\alpha=0.9$ as in Ref. \onlinecite{vanicek:2007}. As the
final results are still fairly close and both agree well with the
experimental value, this demonstrates the relative robustness of the
method once a proper value of $\alpha$ is chosen. In order to achieve
an even higher accuracy and to remove a certain level of arbitrariness
connected with the parameter $\alpha$, one can use, e.g., the distributed
gaussian method \cite{schlegel:2006}. In this method, the EVB potential is fitted
to match the Hessian of an \textit{ab initio} potential at several
points in the transition state region.

\begin{table}
\begin{tabular}{lccccc}
\hline 
$T$ (K) &
QI &
TST &
QI/MP2;GAFF &
exp\footnotemark[1] &
exp\footnotemark[2]\tabularnewline
\hline 
463.25 &
5.5 &
2.7 &
5.4 &
- &
5.3\tabularnewline
468.25 &
- &
2.7&
-&
5.2 &
-\tabularnewline
470.0 &
5.3 &
2.7 &
5.3 &
- &
5.2\tabularnewline
473.15 &
- &
2.6&
-&
5.0 &
-\tabularnewline
478.45 &
5.1 &
2.6 &
5.1 &
5.1 &
5.0\tabularnewline
500.0 &
4.7 &
2.5 &
4.3 &
- &
4.7\tabularnewline
\hline
\end{tabular}

\footnotetext[1]{Ref. \onlinecite{Roth1966}}

\footnotetext[2]{Ref. \onlinecite{Roth1966}}

\caption{The KIE on the first {[}1,5] hydrogen shift reaction 
from Fig.~\ref{fig:equlibrium_pent}. The displayed KIE is defined 
precisely in Eq. (\ref{our_KIE}) of the text.
The column denoted by exp$^{\text{a}}$ contains the raw experimental data from Ref. \onlinecite{Roth1966} whereas the column exp$^{\text{b}}$ contains the data computed from an Arrhenius fit in Ref. \onlinecite{Roth1966}. 
{\footnotesize \label{tab:KIE}}}
\end{table}

\subsection{Temperature dependence of the rate constant}

In contrast to the KIE which is virtually independent of the barrier
height, the rate constant depends on the barrier height exponentially.
Very accurate value of the barrier height is therefore essential for the
calculation of this quantity or its temperature dependence. Table
\ref{tab:therm_dep} contains barrier heights, relative energies of
trans and gauche isomers of (3\textit{Z})-penta-1,3-diene, and TST
values of the rate constant $k_{d_{2}}^{\mathrm{TST}}$ for the
\textbf{1-}\textbf{\footnotesize 1,1-}\textbf{\textit{d}}\textbf{$_{\text{2}}$}
$\rightarrow$ \textbf{1-}\textbf{\footnotesize 5,5-}\textbf{\textit{d}}\textbf{$_{\text{2}}$},
hydrogen shift computed at three different levels of theory using B98/6-311+(2df,p)
optimized geometries. As can be seen from the table, the B98/6-311+(2df,p)
barrier is the lowest one, in accordance with the well known fact that
barrier heights are usually underestimated by the majority of currently
used density functional methods (including B98) \cite{Zhao2005}. Besides B98, two other 
and generally 
more accurate methods used were the CCSD(T) method with the cc-pVTZ 
basis set and the CCSD method without triples correction
but with the substantially larger aug-cc-pVTZ basis set. The difference
between these two methods is still almost 3 $\mathrm{kcal/mol}$ but
at this point it is hard to decide which method is actually closer
to the real value of the barrier. %
\begin{table*}
\begin{tabular}{lcccc}
\hline 
 &
B98/6-311+(2df,p) &
CCSD(T)/cc-pVTZ &
CCSD/aug-cc-pVTZ&
exp\footnotemark[1]\tabularnewline
\hline 
$\Delta\varepsilon^{\ddagger} \mathrm{(kcal/mol)}$&
37.3&
39.9&
42.8&
\tabularnewline
$\Delta\varepsilon^{g-t}\mathrm{(kcal/mol)}$&
3.6 &
3.0&
2.0&
\tabularnewline
$k_{d_{2}}^{\mathrm{TST}}\cdot10^{-6}(s^{-1})[478.45\mathrm{\: K}]$ &
364.6 &
27.8 &
2.8&
7.1\tabularnewline
\hline
\end{tabular}

\footnotetext[1]{Ref. \onlinecite{Roth1966}}

\caption{The barrier height $\Delta\varepsilon^{\ddagger}=\varepsilon^{\ddagger}-\varepsilon_{r}^{\mathrm{trans}}$
measured from the trans conformer (3\textit{Z})-penta-1,3-diene which
is the global minimum, the energy difference 
$\Delta\varepsilon^{g-t}=\varepsilon_{r}^{\mathrm{gauche}}-\varepsilon_{r}^{\mathrm{trans}}$ between trans and gauche conformers,
and the TST rate constant $k_{d_{2}}^{\mathrm{TST}}$ of
the {[}1,5] hydrogen shift reaction 
\textbf{1-}\textbf{\footnotesize 1,1-}\textbf{\textit{\footnotesize d}}\textbf{\footnotesize $_{\text{2}}$} 
$\rightarrow$
 \textbf{1-}\textbf{\footnotesize 5,5-}\textbf{\textit{\footnotesize d}}\textbf{\footnotesize $_{\text{2}}$.}
{\footnotesize \label{tab:therm_dep}}}
\end{table*}

The TST rate constants listed in Table \ref{tab:therm_dep} were computed
using the B98/6-311+(2df,p) partition functions in the HA together
with the barrier height of the method considered.
Neglecting the effect of barrier recrossing and the anharmonicity of the potential energy surface, the TST rate constant should be smaller than the experimental one, because the TST ignores the tunneling contribution that is quite important in this reaction (as demonstrated by the large value of the KIE).
As a consequence, comparison of $k_{d_{2}}^{\mathrm{TST}}$
with the experimental rate constant suggests that both the CCSD(T)/cc-pVTZ 
and especially the B98/6-311+(2df,p) methods give too low barriers. 
The TST rate constant computed with the CCSD/aug-cc-pVTZ
barrier, on the other hand, agrees quite well with the theoretical
expectations.

\begin{figure}
\includegraphics[width=\linewidth]{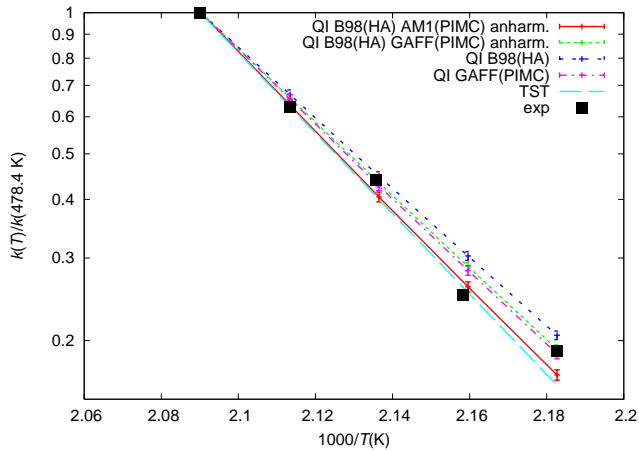}
\caption{Temperature dependence of the rate constant of the {[}1,5] hydrogen shift
reaction \textbf{1-}\textbf{\footnotesize 1,1-}\textbf{\textit{\footnotesize d}}\textbf{\footnotesize $_{\text{2}}$}
$ \rightarrow $
\textbf{1-}\textbf{\footnotesize 5,5-}\textbf{\textit{\footnotesize d}}\textbf{\footnotesize $_{\text{2}}$}
using the CCSD/aug-cc-pVTZ barrier height. The $y$ axis is plotted in
the logarithmic scale. The description of methods used can be found
in Computational details. {\footnotesize \label{fig:therm_dep_H_CCSD}}}
\end{figure}

The QI method provides several theoretical improvements over the TST: Most 
importantly, unlike the TST, the QI approximation includes tunneling and 
other quantum effects. Moreover, its PI implementation
facilitates the calculation
of the anharmonicity effects on the partition functions and delta-delta 
correlation functions. Methods used
to include the anharmonicity correction to the B98/6-311+(2df,p) harmonic
ratios are described in the Computational details. The temperature
dependence of the QI rate constant with the CCSD/aug-cc-pVTZ barrier
is shown in Figure \ref{fig:therm_dep_H_CCSD}. 

{\footnotesize }%
{\footnotesize \par}

As can be seen in Figure \ref{fig:therm_dep_H_CCSD}, both the QI and
TST results agree well with the relatively broad and imprecise
experimental temperature dependence.  Neglecting again the recrossing and 
anharmonicity effects,
one can expect that the value of $k_{d_{2}}^{\mathrm{TST}}$ will
be smaller than the value of $k_{d_{2}}^{\mathrm{QI}}$,
mainly due to the tunneling contribution to the rate that is
captured by $k_{d_{2}}^{\mathrm{QI}}$ but ignored in $k_{d_{2}}^{\mathrm{TST}}$.
Moreover, as the relative importance of the tunneling increases
with the decreasing temperature, the temperature
dependence of the $k_{d_{2}}^{\mathrm{QI}}$ should be weaker than
the dependence of $k_{d_{2}}^{\mathrm{TST}}$. Figure \ref{fig:therm_dep_H_CCSD}
shows that this is indeed true. (The temperature dependence of the relative 
importance of tunneling and of other nuclear quantum effects can be sometimes
more complicated due to other effects such as the temperature dependence 
of the distance between the donor and acceptor \cite{Liu2007}.)

Another interesting feature visible
in the graph is the importance of the anharmonicity correction, which
shows that the change in the temperature dependence caused by the
anharmonicity can be equivalent to the change of the reaction barrier
by several $\mathrm{kcal/mol}$. It is reassuring that the more accurate
AM1 correction influences the result more than the correction computed
using the GAFF force field. This observation agrees well with the expectations based
on the fact that in the GAFF force field the bond stretching coordinates
are actually harmonic and the anharmonicity is caused only by the
other terms of the force field. Nevertheless, both corrections decrease 
the HA free energy of nuclear motion through the lowering
of the ZPE. Finally, the curves of the relative
rate constant in the Arrhenius plot are almost exactly straight lines,
irrespective of the theoretical method used. This suggests that it
would be very hard to observe any non-Arrhenius behavior in this small temperature
range even with a far more precise experimental setup. 

\begin{figure}
\includegraphics[width=\linewidth]{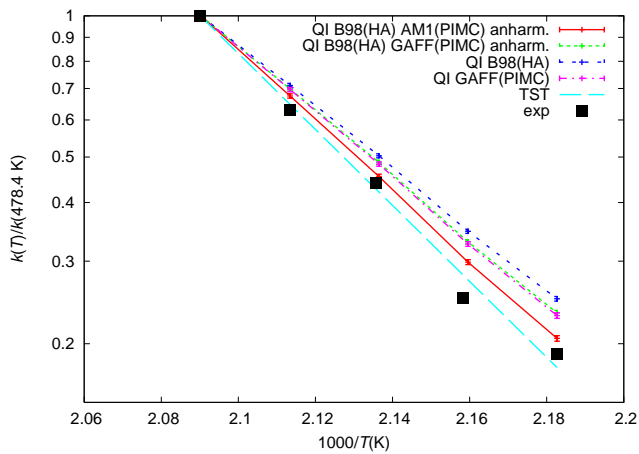}

\caption{Temperature dependence of the rate constant of the {[}1,5] hydrogen
shift reaction \textbf{1-}\textbf{\footnotesize 1,1-}\textbf{\textit{\footnotesize d}}\textbf{\footnotesize $_{\text{2}}$}
$ \rightarrow $
\textbf{1-}\textbf{\footnotesize 5,5-}\textbf{\textit{\footnotesize d}}\textbf{\footnotesize $_{\text{2}}$}
using the CCSD(T)/cc-pVTZ barrier height. The $y$ axis is plotted in
the logarithmic scale. The description of methods used can be found
in Computational details. {\footnotesize \label{fig:therm_dep_H_CCSDT}}}
\end{figure}

The temperature dependence of $k_{d_{2}}$ computed with the CCSD(T)/cc-pVTZ
barrier height is shown in Figure \ref{fig:therm_dep_H_CCSDT}. With
this barrier, the temperature dependence of the more accurate QI method
is only very slightly weaker than the experimental dependence, whereas
the TST rate dependence is still within the range of the experimental
values. However, this apparently excellent agreement of the conventional 
TST with the experiment is most 
likely due to the cancellation of the tunneling and anharmonicity corrections. 
All the other characteristics of the graph are the
same as with the CCSD/aug-cc-pVTZ barrier. When the lowest and probably
the least accurate B98/6-311+(2df,p) barrier is used, curves for all
methods stay above the experimental data in the Arrhenius plot. 

{\footnotesize }%
{\footnotesize \par}

\section{Conclusions}

We have applied a general path integral methodology for computing the EIE, KIE,
and the temperature dependence of the rate constant to the [1,5] sigmatropic
hydrogen shift in pentadiene. In case of the EIE, the computed result has a
higher precision than the experiment \cite{Roth1966}, therefore our result
can be considered as a prediction of the deviation of the exact EIE from the
symmetry determined EIE. In case of the KIE, the accuracy of the previous
result \cite{vanicek:2007} was improved by using a combination of a high level
electronic structure calculation within the harmonic approximation with a path
integral anharmonicity correction using a lower level method. Finally, the
result for the temperature dependence of the rate constant is the first
application of the methodology from Ref. \onlinecite{buchowiecki:2010} to a molecule
with more than three atoms. As for the KIE, the temperature dependence
confirms the importance of tunneling and anharmonicity effects. However,
unlike for the KIE, the accuracy of the energy barrier plays an important role
and therefore a high level method is required. According to our results, 
the CCSD/aug-cc-pVDZ barrier height of 42.8 kcal/mol seems to be the most accurate.

While we have considered a gas phase molecule with 13 atoms, significantly
larger systems can be treated by the presented PI\ approach since only the
light atoms participating in the reaction require a large number $P$ of
replicas. Heavier atoms or atoms less important in a given problem (e.g.,
atoms far from the active site of an enzyme) can be treated either with a
small value of $P$ or fully classically (i.e., with $P=1$). The efficiency
will then be dominated by the cost of an accurate potential energy for which
various strategies are known already from classical molecular dynamics. Among
these are the recent implementations (see, e.g., Ref. \onlinecite{gresh:2007}) 
of polarizable force fields \cite{Warshel:1976,Warshel2007} or the accelerated sampling using hybrid
quantum-mechanical/molecular-mechanical potentials \cite{kamerlin:2009}. A
combination of such techniques with the methodology described in this paper
will be the subject of future applications.

\subsection*{Acknowledgments}

This research was supported by the Swiss NSF (Grant No. $200021\_124936/1$)
and by the EPFL. Authors thank M. Buchowiecki for helpful discussions.


\end{document}